\documentclass[prl,twocolumn,preprintnumbers,amsmath,amssymb,
floatfix,showpacs]{revtex4}
\usepackage{hyperref}
\usepackage{amsmath}
\usepackage[dvips]{epsfig}
\newcommand{\beq}{\begin{equation}}
\newcommand{\eeq}{\end{equation}}
\newcommand{\beqa}{\begin{eqnarray}}
\newcommand{\eeqa}{\end{eqnarray}}
\newcommand{\ba}{\begin{array}}
\newcommand{\ea}{\end{array}}

\begin{document}
\title{Kohn-Sham theory of rotating dipolar Fermi gas in two dimensions}

\author{Francesco Ancilotto$^{1,2}$} 
\affiliation{$^1$Dipartimento di Fisica e Astronomia "Galileo Galilei" and CNISM, 
                 Universit\`a di Padova, via Marzolo 8, 35122 Padova, Italy \\
             $^2$CNR-IOM Democritos, via Bonomea, 265 - 34136 Trieste, Italy} 

\begin{abstract} 
A two-dimensional dipolar Fermi gas in harmonic trap 
under rotation is studied 
by solving "ab initio" Kohn-Sham equations.
The physical parameters used match those
of ultracold gas of fermionic $^{23}Na^{40}K$ molecules,
a prototype system of strongly interacting
dipolar quantum matter, which has been created
very recently.
We find that, as the critical rotational 
frequency is approached
and the system collapses into the lowest Landau level,
an array of tightly packed quantum vortices develops, in spite of 
the non-superfluid character of the system.
In this state the system looses axial symmetry, 
and the fermionic cloud boundaries assume 
an almost perfect square shape.
At higher values of the filling factor the
vortex lattice disappears, while the system 
still exhibits square-shaped boundaries.
At lower values of the filling factor
the fermions become instead localized 
in a "Wigner cluster" structure.
\end{abstract} 

\date{}
\maketitle

Recent developments in the field of 
ultracold dipolar atoms have boosted 
investigations of many-body effects 
associated with long-range interactions (see, e.g., the 
review in Ref.\cite{bar08}). 
Recently, the creation of ultracold dipolar gas of fermionic
molecules with large intrinsic dipole moments
has been achieved\cite{kni08,zwi15}, opening the
way to explore the intriguing many-body physics of correlated 
Fermi systems
associated with the long-range, anisotropic nature
of dipolar interaction between molecules\cite{bar12},
which include topological superfluidity\cite{bru08,coo09}, interlayer pairing
between two dimensional systems and the formation of dipolar 
quantum crystals\cite{sie11}.

Two-dimensional (2D) dipolar systems are
particularly interesting, since the lifetime of 
heteronuclear Feshbach molecules with permanent 
electric dipole moment increases by the confinement in 
two dimensions\cite{dem11}. Indeed such polar molecules 
can have very large dipole moments, of the order of 1 Debye, allowing to 
access the regime of strong correlations in a controllable way.
A 2D dipolar Fermi liquid, which is stable at low density, is expected
to convert into a Wigner crystal at high densities\cite{bar08,zyl15}.
For intermediate values of the interaction strengths
an instability at finite wave vector is predicted, driving the
system to a "stripe"-phase\cite{par12,sun10,blo14} 
(see also Ref.\cite{bar08} and references 
therein).
Recent Quantum Monte Carlo study confirmed the liquid-solid transition
at high coupling but
found that the stripe phase is never energetically favored
\cite{gio12}. 

A particular fascinating route towards the realization of
strongly correlated system of ultra-cold gases (either bosonic
or fermionic) is the use
of rapidly rotating harmonic traps. 
When the rotational frequency approaches the trap frequency,
i.e. just below the limit of 
centrifugal instability, the single particle
energy spectrum becomes highly degenerate and hence the 
kinetic energy of the Fermi system is much reduced, thus
enhancing the role of the interparticle interactions.

A uniform rotation with angular velocity 
approaching the centrifugal limit is in fact formally equivalent to 
a magnetic field (in the rotational frame) that 
regroup single-particle states into discrete, highly
degenerate Landau Levels (LL). 
Such equivalence, which holds in 2D and in the presence of an harmonic
trapping potential only,
is embodied in the following formal identity\cite{fetter} involving the 
many-body Hamiltonian of the (interacting) system
in the rotating reference frame: 
 
\begin{eqnarray}
{\it H}=\sum _{i=1}^N (  {{\bf p}_i ^2 \over 2M}
+{M \over 2}\omega _h^2 r_i^2-\Omega \hat {L}_{iz})
+V \nonumber \\
=\sum _{i=1}^N [ {1 \over 2M}({\bf p}_i-M\omega _h 
{\bf e}_z\times {\bf r}_i)^2+{M\over 2}
(\omega_h^2 -\Omega^2) r_i^2]+V \,\,\,
\label{energy}
\end{eqnarray}
where
$V$ is the interaction energy, 
$\Omega $ is the rotation frequency and $\hat {L}_{iz}$ is the projection
of the angular momentum of the $i$-th particle along the $z$ axis.
Here ${\bf r}_i=x_i{\bf e}_x+y_i{\bf e}_y$ is the position vector
of the $i$-th particle. When $\Omega =\omega _h$
the non-interacting part reduces to the Landau Hamiltonian
of particles with mass $M$ and charge $e$ 
moving in a constant magnetic field ${\bf B}=B{\bf e}_z$,
of strength $B=2M\Omega /e$.
The eigenvectors of the non-interacting part span Landau levels 
with energies $\epsilon _n=\hbar \omega _c(n+1/2)$, where $\omega _c=2\Omega$.

We consider in the following a two-dimensional, spin-polarized dipolar Fermi
gas, characterized by an interaction term in Eq.(1)
$
V=\sum _{i<j}^N {d^2 \over |{\bf r}_i-{\bf r}_j|^3}
$
Here $d$ is the magnetic dipole moment of an atom/molecule and 
${\bf r},{\bf r}^\prime$ are coordinates in the 2D $x-y$ plane.
Being the dipole moments aligned parallel to the
z-axis, the (long-range) pair potential is purely {\it repulsive}.
The range of the dipole-dipole interaction 
is characterized by the length
$r_0=Md^2/\hbar ^2$.

Exotic forms of vortex lattices,
e.g. square, stripe- and bubble-"crystal" lattices
are expected in rotating Bose Einstein condensates
when the critical rotational frequency is approached\cite{coo05,sch04,bar05}.
Rotating dipolar Fermi gases have been proposed \cite{lun00,ost07} 
as suitable candidates to realize Laughlin-like state and more exotic 
quantum liquids, as well as their crossover behavior to Wigner crystals.
At variance with the case of a non-rotating
dipolar Fermi gas in a 2D trap, where the crystalline state becomes
energetically favored at high densities, 
in the case of a fast rotating
dipolar gas, the situation is reversed\cite{bar08}:
rotating dipoles in the Lowest LL (LLL) behave similarly to
electrons, where the crystalline phase is stable at 
low densities. 
Indeed, it has been shown\cite{lew08,jhe13} that a rapidly rotating polarized 
2D dipolar Fermi gas undergoes a transition to a 
crystalline state, similar to the two-dimensional
Wigner electron crystal in a magnetic field,
for sufficiently low value of the filling factor, $\nu <1/7$.
Here 
$\nu \equiv 2\pi l^2 n_F$ (where $n_F$ is the areal density of the fermionic system
and $l=\sqrt{\hbar/M\omega _h}$
is the magnetic length)
gives the fraction of the occupied LLL.
At filling factor $\nu =1/3$ the system is instead
well described in terms of fractional quantum Hall-like states\cite{bar05}.

Density Functional Theory
(DFT), which is perhaps the most widely used and successful technique 
in electronic structure calculations of condensed matter
systems, has only recently entered the field of cold gases
as a useful computational tool which goes beyond the mean-field
description by taking into account correlation effects, 
and thus is capable to
yield quite accurate results in agreement with
more microscopic (but also much more computationally expensive) 
approaches.
The well-known Kohn-Sham (KS) mapping\cite{kohn} of the
many-body problem into a non-interacting one make this approach 
applicable in practice, often within the so-called 
Local Density Approximation (LDA)\cite{kohn}.
Recently, KS-DFT has been applied to cold atomic Fermi gases in 
optical lattices\cite{tro12}
and to the study of unitary trapped Bose gas
\cite{anci}. 
DFT approaches have been used recently to describe a Fermi dipolar
system in various "single-orbital" approximations 
(Thomas-Fermi \cite{zyl15}, Thomas-Fermi-Dirac \cite{fan11},
Thomas-Fermi-von Weizsacker \cite{zyl13,zyl14}).
In Ref.\cite{abe14} a parameter-dependent 
DFT-LDA approach was used to study small
number of harmonically trapped fermions.
A somewhat different density functional formalism,
whose applicability is however limited to a small number
of particles, and which is
based on the self-consistent combination of the
weak and the strong coupling limits, has been proposed 
to study the ground-state 
properties of strongly correlated
dipolar and ionic ultracold bosonic and fermionic gases\cite{gor15}.

Here we use the conventional KS approach, based on 
accurate description for the correlation energy 
of the dipolar system as 
provided by Diffusion Monte Carlo calculations\cite{gio12}.
Our approach does not require any adjustable parameter,
and thus belongs to the family of the "ab initio" methods well
known in the electronic structure community.
The Kohn-Sham formulation\cite{kohn} of Density Functional 
Theory \cite{gross} for an inhomogeneous system of $N$ interacting  
particles with mass $M$ is based on the following energy functional of the density
which includes the exact kinetic energy of a fictitious non-interacting
system and the interaction energy functional $E_{HFC}$:

\beq
E_{KS}[\rho ]=-{\hbar ^2 \over 2M}\sum _i \int{\phi _i^\ast({\bf r}) 
\nabla ^2 \phi _i({\bf r})d{\bf r}} +E_{HFC}[\rho ]
\label{k-s}
\eeq

The $\{\phi _i({\bf r}),\,i=1,N\}$ are single-particle orbitals,
forming an orthonormal set, $\langle \phi _i | \phi _j \rangle=\delta _{ij}  $,
filled up to the Fermi level.
The total density of the system is
$\rho ({\bf r})=\sum _{i=1}^N |\phi _i({\bf r})|^2$

$E_{HFC}$ is the sum of the
direct+exchange dipolar interaction term 
(usually termed "Hartree-Fock" energy, $E_{HF}$)
and the correlation energy ($E_C$).
The Hartree-Fock energy of a dipolar Fermi gas in 2D 
has two contribution. The first,
for the homogeneous system of surface density $\rho =N/A$, is:

\beq
 \label{hf1}
E_{HF}^{(1)}= {256\over 45 }N d^2 \sqrt{\pi }\rho ^{3/2} 
\eeq

The second term is {\it non-local} in nature 
and is given by\cite{fan11,zyl13}

\beq
 \label{nloc}
E_{HF}^{(2)}= -\pi d^2 \int d {\bf r} \rho ({\bf r}) 
\int d {\bf r}^\prime 
\int {d {\bf k} \over (2\pi)^2} k e^{-i{\bf k}\cdot 
({\bf r}-{\bf r}^\prime)}\rho ({\bf r}^\prime)
\eeq

This term vanishes in the uniform limit, while the negative sign
crucially lower the total energy of the system in inhomogeneous 
configurations. This term has been shown to be 
essential to stabilize structures such as one-dimensional stripe phases
and the Wigner crystal that is expected at high densities \cite{zyl15}.

In the following we will treat $E_{HF}^{(1)}$ and $E_C$ within the
Local Density Approximation (LDA), i.e.

\beq
 \label{lda}
E_{HF}^{(1)}+E_C=\int  
[ {256\over 45 } d^2 \sqrt{\pi }\rho ({\bf r})^{5/2}+
\rho ({\bf r}) \epsilon _C(\rho ({\bf r}))] d{\bf r}
\eeq
where $\epsilon _C (\rho )$ is the correlation energy per particle 
of the {\it homogeneous} system of density $\rho $, as obtained
from the (virtually exact) Diffusion Monte Carlo
calculations of Ref.\cite{gio12}.

The total energy functional in the co-rotating frame with constant 
angular velocity $\Omega $ 
(where the dipolar system appears at rest) and in the
presence of an isotropic harmonic trapping potential
of frequency $\omega _h$,
$U({\bf r})=\frac{1}{2}M\omega_h^2(x^2+y^2)$, is given by:

\beq 
 \label{dft}
 E[\rho ] = E_{KS}[\rho ] + \int d{\bf r} \rho ({\bf r})U({\bf r}) -\Omega \langle L_z\rangle
\eeq
Here $\langle L_z\rangle$ is the total angular momentum of the system.
Constrained minimization of the above functional
leads to the coupled KS eigenvalues equations
\beq
 \label{kseq}
  [-{\hbar ^2 \over 2M }\nabla ^2 + V_{KS}]\phi _i({\bf r})=\epsilon _i \phi _n({\bf r})
\eeq
where 
\begin{eqnarray}
 \label{ham}
  V_{KS}({\bf r})=\epsilon _C(\rho ({\bf r}))+\rho ({\bf r})
{\partial \epsilon _C \over \partial \rho}
+{128\over 9}d^2\sqrt{\pi}\rho ^{3/2}({\bf r})
-\Omega \hat {L}_z \nonumber \\
-2\pi d^2 
\int d {\bf r}^\prime 
\int {d {\bf k} \over (2\pi)^2} k e^{-i{\bf k}\cdot 
({\bf r}-{\bf r}^\prime)}\rho ({\bf r}^\prime) \,\,\,\,\,
\end{eqnarray}
and $L_z=-i\hbar (x\partial /\partial y-y\partial /\partial x)$.

We seek for
stationary solutions $\{ \phi _i ({\bf r}),i=1,N \}$ 
by propagating in imaginary time the time-dependent version\cite{kohn} of 
the KS equations (\ref{kseq}).
Both the density and the orbitals $\phi _i$ have been
discretized in cartesian coordinates using
a spatial grid fine enough to guarantee
well converged values of the total energy. 
The orthogonality between different orbitals has been enforced
by a Gram-Schmidt process. The spatial
derivatives entering Eq.(\ref{kseq}) have been calculated with 
accurate 13-point formulas, while
Fast-Fourier techniques have been used to efficiently
calculate the non-local term entering the KS potential $V_{KS}$.

We take in our calculations $d=0.8$ Debye, which is 
appropriate to $K_{40}Na_{23}$ molecules in
the experimental realization of Ref.\cite{zwi15}.
The mass is that of a $K_{40}Na_{23}$ molecule.
The range of the potential is thus 
$r_0=Md^2/\hbar ^2\sim 0.6\,\mu m \sim 0.2\,a_H$, $a_H=\sqrt{\hbar/2 M \omega _h}$ 
being the oscillator length.
In the ground-state of the non-rotating system, 
the adimensional interaction strength characterizing the system
is $k_F r_0\sim 0.9$ (where $k_F=\sqrt{4\pi \rho _{max}}$ 
is the Fermi wavevector of the 2D system at a density
equal to the maximum density in the center of the trap), i.e.
a relatively weak value which can easily be achieved in 
experiments. The interparticle distance $\langle r\rangle$ 
is larger than the range of the interaction, being $\langle r\rangle/r_0\sim 3.6$.
The corresponding dipolar interaction
energy $E_d=d^2/\langle r\rangle ^3$ approaches 20\% of the
local Fermi energy $\hbar ^2 k_F^2/2M$.
In spite of the relatively weak coupling,
as a consequence of the rotation,
strong correlation effects will show up in the
density distribution of the calculated stationary states,
as shown in the following.
We consider systems with up to $N=200$ fermions.

\begin{figure}
 \epsfig{file=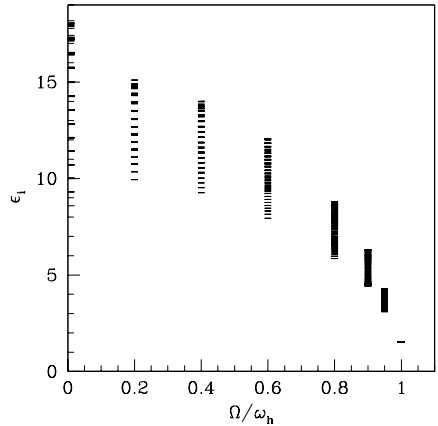,width=8 cm,clip=}
 \caption{Calculated KS eigenvalues for $N=100$ fermions, 
for different values of the rotational frequency $\Omega $ }
 \label{fig1}
\end{figure}

We show in Fig.\ref{fig1} the evolution of the calculated single particle 
KS eigenvalues $\epsilon _i$ for the case $N=100$, 
as the rotational frequency $\Omega $ 
approaches from below the harmonic frequency $\omega _h$.
At $\Omega = 0.999\,\omega _h$ it appears that all the 
energy levels collapse into a single level, the 
(highly degenerate) LLL.

\begin{figure}
 \epsfig{file=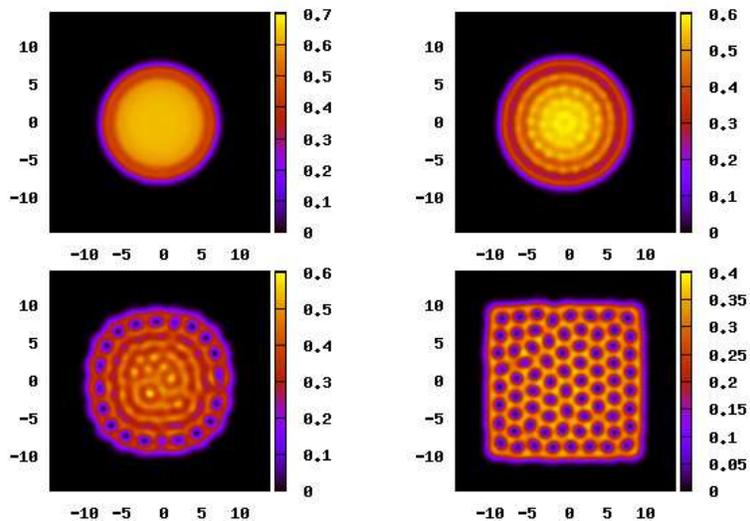,width=10 cm,clip=}
 \caption{(Colors online) 
Density of the dipolar system ($N=100$) at selected values of 
$\Omega /\omega_h$. From left to right and from top to bottom:           
$\Omega /\omega_h=0.96,\,0.98,\,0.99,\,0.999$.
The x-axis coordinates are in units of $a_h$, while the
density is in units of $a_h ^{-2}$.}
 \label{fig2}
\end{figure}

It is instructive to follow how the density
of the system evolves as $\Omega $ is increased.
This is shown in Fig.\ref{fig2}, where the densities of 
selected configurations corresponding
to different values of $\Omega $ are displayed.

For low values of $\Omega $ ($\Omega =0$ included) the calculated stationary 
states have circular symmetry, and the density has the familiar, almost featureless 
shape of 
a trapped cold gas cloud. As soon as $\Omega $ approaches $\omega _h$, however,
a ring of equally spaced deep dimples develop close to the periphery of the
cloud, while the systems looses its axi-symmetric shape.
Eventually, very close to $\omega _h$, an array of tightly packed 
vortices develops,
similarly to the Abrikosov lattice of vortices in 
rotating superfluids, while the system boundaries acquire 
a surprising {\it square} shape.


The calculated current density in the 
state with $\Omega /\omega_h=0.999$ in Fig.\ref{fig2}
appears indeed to be 
circulating around the zero-density minima (black dots in 
the last panel of Fig.\ref{fig2}),
as expected for a vortex array.
The total angular momentum $\langle \hat {L}_z\rangle $ shows also the typical
behavior associated to the nucleation of 
quantum vortices, i.e. a sequence of rounded steps 
(with amplitudes $\sim N\hbar $) with increasing 
rotation frequency $\Omega $, as more vortices are nucleated in the
system during the minimization process leading to the stationary 
state shown in the last panel of Fig.\ref{fig2}.
The average distance $d_v$ between vortices in the structure 
shown in Fig.\ref{fig2})
is $\sim \,20\%$ larger than the one calculated (assuming a triangular
vortex lattice of areal density $n_v$) using 
Feynman's formula\cite{Fey55}, $n_v=2/\sqrt{3}d_v^2=M\Omega /\pi \hbar$.

Quantized vortices in Fermionic cold gases are usually 
associated to pairing interactions, as in the 
BCS side of a unitary Fermi gas\cite{zwi05}, where they 
are considered the hallmark of the superfluid character of the system.
The presence of vortices in a system with purely repulsive
interactions, like the one studied here, 
has been predicted to occur in fermion systems
with purely repulsive interaction such as quantum dots,
where the rotation is induced by an external magnetic field
(see for instance Ref.\onlinecite{saa04}). 
Indirect evidence of 
vortices in ultra small fermion droplets ($N=6$) with
aligned dipoles 
has been provided in Ref.\onlinecite{eri12}. 
However, due to the implicit symmetry constraints in 
the calculations of Ref.\onlinecite{eri12} multi-vortex structures
like the one shown in Fig.\ref{fig2}
did not show up in the calculated density profiles. 

A striking feature of the $N=100$ system in the LLL 
is the lack of axial symmetry represented by the 
unusual square-shaped boundaries.
This seems to be intimately connected with 
the interactions between fermions: 
the system shown in Fig.\ref{fig2}, under the same conditions but 
with no interactions between fermions,
exhibits density profiles with circular symmetry 
all the way up to $\omega_h$.
Deviations from axi-symmetric configurations in 
isotropic trapping have been found in fast rotating BEC
at overcritical rotation\cite{rec01}, as a consequence of
the interatomic forces.
Stable, non axi-symmetric multi-lobed shapes
also characterize the fast rotation of classical liquid droplets\cite{bro80}.

\begin{figure}
 \epsfig{file=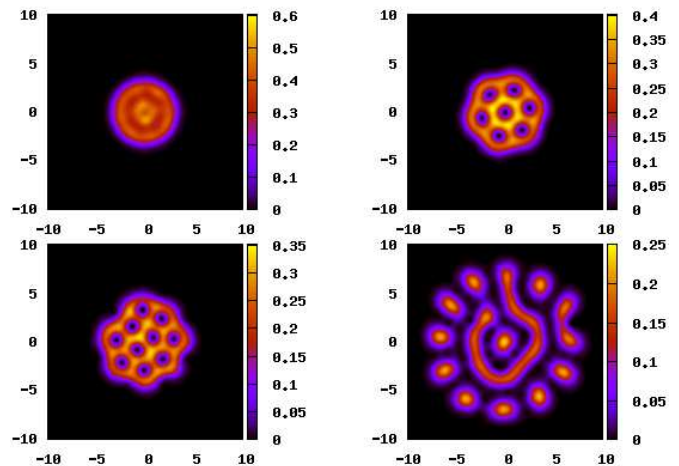,width=9 cm,clip=}
 \caption{(Colors online)  
Density of the dipolar system ($N=13$) at selected values of 
$\Omega /\omega_h$ From left to right and from top to bottom:           
$\Omega /\omega_h=0.90,\,0.96,\,0.98,\,0.999$.
The x-axis coordinates are in units of $a_h$, while the
density is in units of $a_h ^{-2}$.
          }
 \label{fig3}
\end{figure}

The configuration shown in the lowest panel of Fig.\ref{fig2}
corresponds to a filling factor $\nu \sim 0.77$.
By decreasing the number of fermions in the trap we can reach
lower values of $\nu $. One example is 
shown in Fig.\ref{fig3}, where $N=13$.
Again, as the centrifugal limit is approached,
stationary configuration with an increasing number of vortices are found:
vortices enter the fermion droplet from the low density periphery (a mechanism 
common to BEC\cite{but99} and Helium-4\cite{anc14}).
As $\Omega \sim \omega_h $ (last panel in Fig.\ref{fig3}), however,
a completely different pattern shows up, resembling a cluster of
localized particles (albeit with a partially melted second shell).
This configuration is characterized by $\nu \sim 0.18$.
We take this as a clear evidence of formation
of a "Wigner cluster" structure for sufficiently low
values of the filling factor.

\begin{figure}
 \epsfig{file=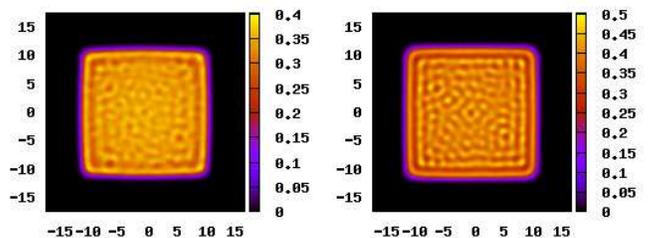,width=9 cm,clip=}
 \caption{(Colors online) 
Density of the dipolar system at $\Omega /\omega_h=0.999$
for $N=160$ (left panel) and $N=200$ (right panel), corresponding
to filling factors $\nu=1.08$ and $\nu=1.25$, respectively.
          }
 \label{fig4}
\end{figure}

Higher values of $\nu $ can be conversely achieved by increasing
the fermions number. In this case the vortex lattice disappears,
and the smoother structures shown in Fig.\ref{fig4} develop.
Here, $\nu=1.08$ and $\nu=1.25$, respectively (corresponding
to $N=160$ and $N=200$ fermions).
Note however that the peculiar square-shaped boundaries  
remain even at higher values of the filling factor.

Although small numbers of cold trapped atoms, like
the ones considered here, can nowadays be
achieved in experiments\cite{ser11}, 
we expect that the surprising phenomenology
revealed by our calculations should be present
also in larger systems of rotating 
dipolar fermionic molecules in quasi 2D
harmonic traps. 
Achieving rotation frequencies $\Omega \sim 0.999\,\omega_h $
is a challenging task, but definitely within the reach of 
current experiments\cite{fetter}.
Due to the relatively high contrast of the vortex array
shown in Fig.\ref{fig2}, its observation  
should be possible by direct imaging
of the atomic cloud after expansion (the vortex lattice in quantum
dots still awaits experimental detection).




The author thanks S.Giorgini and N.Matveeva for having 
shared their DMC numerical results, and
S.Giorgini, L.Salasnich, F.Toigo, J.Boronat and M.W.Cole for useful discussions
and comments.

\end{document}